\documentclass[prl,aps,twocolumn,showpacs,floatfix]{revtex4}
\usepackage{graphicx}
\usepackage{bm}
\usepackage{amsmath}
\usepackage{amssymb}
\usepackage{amsfonts}
\usepackage{color}
\usepackage {ulem}
\usepackage{dcolumn}%
\usepackage{bm}

\newcommand{\be}{\begin{equation}}
\newcommand{\ee}{\end{equation}}
\newcommand{\bea}{\begin{eqnarray}}
\newcommand{\eea}{\end{eqnarray}}

\begin{document}
\title{Creep and fluidity of a real granular packing near jamming}
\author{Van Bau Nguyen$^{1,2}$, Thierry~Darnige$^{1}$, Ary~Bruand$^{2}$, Eric~Clement$^{1}$}
\affiliation{$^{1}$PMMH, ESPCI, CNRS (UMR 7636) and Univ. Paris 6 \& Paris 7, 75005 Paris France.\\
$^{2}$CNRS/INSU, ISTO (UMR 6113), Univ. Orleans, 45071 Orleans France.} 
\date{\today}
\begin{abstract}
We study the internal dynamical processes taking place in a granular packing below yield stress.
At all packing fractions and down to vanishingly low applied shear, a logarithmic creep is observed. 
The experiments are analyzed using a visco-elastic model which introduces an internal, time dependent, "fluidity" variable.
For all experiments, the creep dynamics can be rescaled onto a unique curve 
which displays jamming at the random-close-packing limit.
At each packing fraction, we measure a stress corresponding to the onset of internal granular reorganisation and
a slowing down of the creep dynamics before the final yield. 
\end{abstract}
\maketitle
\begin{figure}[htb]
\begin{center}
\vspace{-0.9cm}
\includegraphics[width=0.48\textwidth]{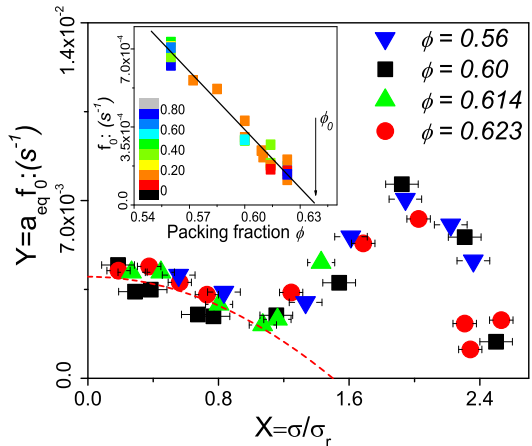}
\vspace{-1cm}
\caption{Rescaled ageing parameter ${a_{eq}}{f_0}$ as function of rescaled of shear stress $\frac{\sigma }{{{\sigma _r}}}$, dotted line: $y=\frac{1}{{{t_*}}}\left( {1 - {{(\frac{x}{{1.5}})}^2}} \right)$ with $t_* \cong 250 s$. Inset : initial fluidity $f_{0}$ as a function of packing fraction for different shear stresses, the color index reflects the ratio $\sigma/ \sigma_M$, fit line: $y = F({\phi _0} - \phi )$ with $F=0.0086$ and ${\phi _0}=0.635 \pm 0.002$.}
\vspace{-0.7cm}
\label{collapse} 
\end{center}
\end{figure}

For granular matter, it is currently accepted that a quasi-static limit exists as for grains of macroscopic size, thermally activated processes can be ignored. At low shear rate, mechanical properties of granular packing are usually described by rate independant constitutive relations \cite{Wood_1990}. However, there are compelling experimental evidences that this limit is just a short-time approximation and time dependent processes are significant on the long run \cite{Schmertmann_1991}. Many numerical simulations based on soft interparticular interactions \cite{Ohern_2003} have brought to the front the idea of an "universal" jamming transition scenario based on a mechanical rigidity threshold separating solid and fluid behavior (see a recent review and refs in (\cite{Hecke_2010})). However for real grains, interparticle solid friction was shown to affect the rigidity onset and stabilize packing at compacity below the random close packing limit \cite{Hecke_2007}. In this case, experiments have pointed out the central importance of nanometric scales where humidity \cite{Bocquet_1998}, contact plasticity, tiny thermal variations\cite{Chen_2006} or weak mechanical external noise\cite{Caballero_2009}, do impact significantly the macroscopic dynamics and the rheology. Since the original theoretical propositions of "soft glassy rheology" or "`Shear Transformation Zones" \cite{Sollich_1997} various models have tried to capture the complex energy reorganization dynamics taking place in amorphous solids or yield-stress fluids, in relation with their constitutive rheological laws. Microscopically, the emergence of plasticity is often explained via a simple picture where localized elastic instabilities release irreversibly long range elastic constraints \cite{Picard_2005} which may be organized spatially as shear driven avalanches\cite{Lemaitre_2009}. Macroscopically, to account for the complex phenomenology, an internal time-dependent variable called "fluidity" is often introduced \cite{Derec2001} to describe the rate of stress relaxation.
%
%
%
\begin{figure}[t!]
\includegraphics[width=\columnwidth]{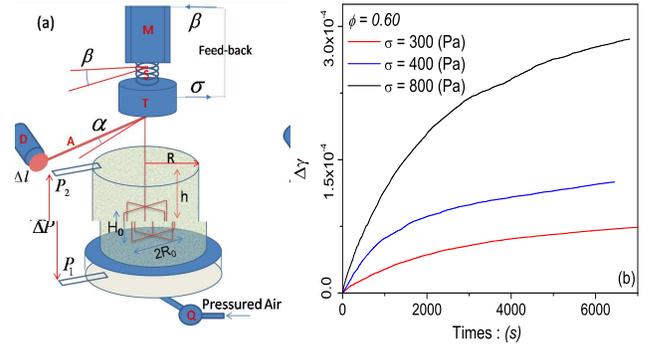}
\caption{(a)Schematics of the shear cell. (M): motor, (S): torsion spring, (T): torque probe, (D): induction distance probe, (A): transversal arm, (Q): flowmeter, ${P_1}$: differential pressure probes. (b) Creep deformation : $\Delta \gamma (t)$ under constant shear at packing fraction $\phi  = 0.60$.}
\vspace{-1cm}
\label{setup}
\end{figure}
In this paper, we study the creeping dynamics of granular packing under constant shear stres, below the Coulomb limit. In order to reveal the internal relaxation processes and make connection with the behavior of a large class of yield stress fluids \cite{Caton_2010}, we propose a quantitative analysis using a  phenomenological model based on the "fluidity" concept.\\
All the mechanical tests are performed at well defined packing fractions $\phi$ (see fig.\ref{setup}). To achieve this goal, the set-up is designed as an air fluidized bed. The container is a plastic cylinder of inner diameter $D=10cm$ closed at its bottom by a metal grid supported by a honeycomb grid. Pressurized air is introduced in an admission chamber below the grid at a controlled over-pressure $\Delta P$. We use glass beads of density  $\rho=2500kg/m^{3}$ and mean diameter $d=200 \mu m$ (R.M.S. polydispersity $\Delta d = 30 \mu m$). A mass $M$ of grains is poured into the container such that the typical packing height is $L=10cm$.  Using a flow rate just above the fluidization value, we obtain after stoppage, an initially loose granular packing at a compacity $\phi \approx 0.56$. Then, by successive tapping on the container side, the packing fraction can be increased up to the desired packing fraction (maximal value $\phi=0.625$). Note that in  the present report, the relative humidity is kept at $35 \pm5\%$ and we work in a "quiet" environment characterized by a backgound noise quantified by placing in the packing an accelerometer (RMS acceleration $0.02 ms^{-2}$ ).
\begin{figure*}[htb]
\begin{center}
\includegraphics[width=0.95\textwidth]{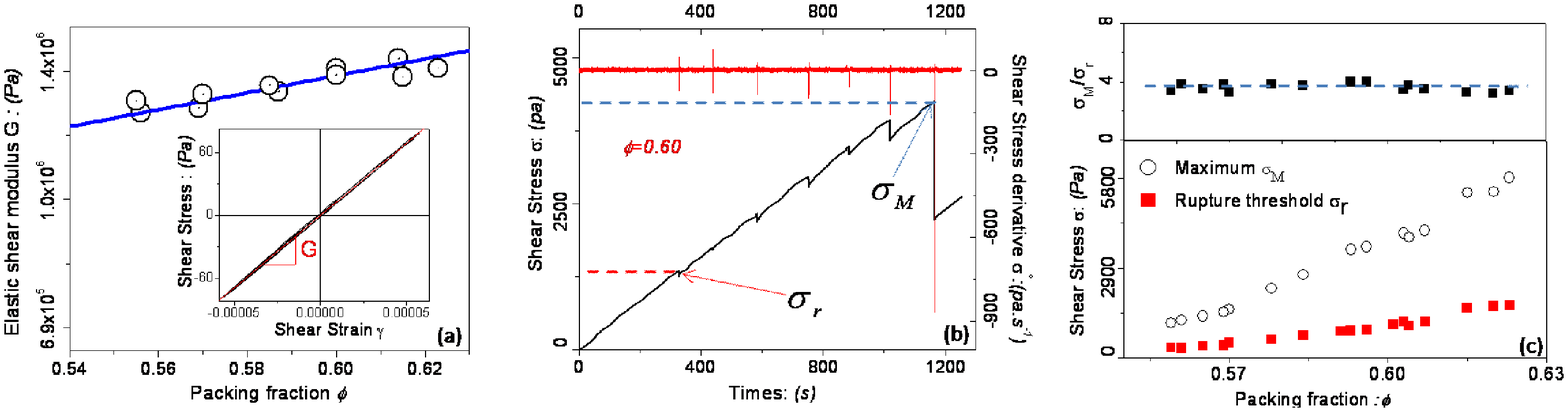}
\caption{. (a) Shear modulus $G(\phi)$ under gravity confinement. Straight line $y=G_0 x$, with $G_0=2,28.10^{6}~Pa$. (b) Response to a stress ramp : shear stress $\sigma$ as a function time for $\mathop \beta \limits^{.}=0.00104~rd/s$ and a packing fraction $\phi=0.6$. First rupture stress $\sigma_r$ and maximal stress $\sigma_M$ are displayed as horizontal dotted lines. (c) For the same rotation rate, $\sigma_r$, $\sigma_M$ and the ratio $\frac{{{\sigma _M}}}{{{\sigma _r}}}$ as a function of packing fraction.}
\vspace*{-0.8cm}
\label{ramp} 
\end{center}
\end{figure*}	
Before each mechanical measurement, the packing fraction value is evaluated from a linear fit between pressure drop $\Delta P$ and flow rate $Q$ : $\Delta P =\lambda Q $, stemming from Darcy's law \cite {Bear89}. The relation between permeability $K$ and packing fraction was calibrated by preliminar series of experiments and a Carman-Kozeny relation was obtained : $K(\phi)=A\frac{(1-\phi)^{3}}{\phi^{2}}d^{2}$, with $A=1/165$. Consequently, for a mass $M$ of grains poured in the cylinder, the packing fraction is obtained through the relation : $K(\phi)~\phi  = \eta \frac{{4~M}}{{\lambda \pi {D^2}\rho }}$, where $\eta=1,85 x 10-6~Pa.s$ is the air viscosity.
To shear the granular packing, we use a four-blade vane in stainless steel of height $H_0=2.54cm$ and diameter $2R_0=2.54cm$ (see fig.\ref{setup}a),introduced at a depth $h=5cm$ below the surface prior to the initial fluidization process. This procedure creates reproducible initial conditions. Shear stress is applied through the vane (see fig.\ref{setup}) connected axially to a torque probe \textit{(T)} itself coupled to a brushless motor \textit{(M)} via a torsion spring \textit{(S)}. The vane rotation angle $\alpha$ is monitored via a transversal arm \textit{(A)} whose rotation is followed by a displacement induction probe \textit{(D)}. The motor rotation angle $\beta$ is imposed with a $2 \pi /10000$ precision. Torque and displacement signals as well as the motor command are connected to a Labview controler board. The last one is programmed to impose a motor rotation rate or a fixed stress using a feedback loop on the torque signal. 
In the following, we ignore the stress and strain spatial distribution due to the Couette cell geometry and define only average values obtained from mesurements of angular rotation $\alpha$ and torque $T$. The mean packing deformation $\gamma$ is defined as $\gamma  = \frac{\alpha R_0}{R-R_0}$ and the mean shear stress is $\sigma =\frac{T}{2\pi R_{0}^{2}H_{0}}$. On fig.\ref{setup}(b), we display three examples of creep curves $\Delta \gamma(t)=\gamma(t)-\gamma(0)$ obtained at fixed compacity and shear stress values $\sigma$.\\ 
\textsl{Elastic response~}- To obtain the elastic response of the packing initially prepared at a given packing fraction, stress cycles were performed corresponding to sinusoidal deformations of small amplitudes around $10^{-5}$. The cycles were carried out under constant mean confining pressure (hydrostatic loading). In the short time limit, the response is essentially reversible (see inset of fig.(\ref{ramp}a). The effective elastic shear modulus increases with packing fraction almost linearly : $G_{eff}=G_{0} \phi$ (see line on fig.\ref{ramp}a), with a value of $G_{0}= 2,28.10^6 Pa$. Interestingly this simple result is consistent with a mean-field Hertz elasticity theory (see\cite{Makse_2005} and refs inside) where under a confining pressure $P_0= \rho \phi g h$, the shear modulus scales as $G_{eff} \propto E_{0}\left( \phi Z\right) ^{2/3}\left( \frac{P_0}{E_{0}}\right) ^{1/3}$, where $Z$ is a constant mean contact number and $E_0$ the material Young's modulus, meaning that, in the range of density explored, the contact density at the origin of the $\phi Z$ term varies linearly with $\phi$ .\\
\textsl{Response to a stress ramp~}- To identify the the maximal stress supported by the packing before yield, shear stress ramps were applied at a constant motor rotation rate ($\mathop \beta \limits^{.}$), using the softest spring constant available ($k=2,45.10^{-3}Nm/rd$). On figure \ref{ramp}(a), the stress response at $\phi=0.60$ is displayed as a function of time. At first, a linear increase of the stress is observed with a slope corresponding to the spring constant. Then, at a given stress level $\sigma_{r}$, we observe the emergence of well-marked and sudden granular material reorganizations (see top-view on fig.\ref{ramp}(b)) in the form of rather equidistant events corresponding to stress drops and large plastic deformations ($\delta \gamma=10^{-3}-10^{-2}$). We define this stress value as the "first rupture stress" : $\sigma_r$. However, stress can still be increased but undergoes series of partial rupture, up to a maximal value $\sigma_M$ followed by a large stress jump. Then, a subsequent stick-slip dynamics is observed. Interestingly, such a structured fluctuation regime was also observed by Albert et al.\cite{Albert_2000} for the drag force on an intruder near jamming. In our case, the maximal stress value corresponds to a Coulomb yield criterion as we verified that its value increases linearly with the confining pressure. Such experiments were performed varying packing fraction and ramp velocities and we only display here stresses obtained at the slowest driving velocity where the values are quite insensitive to the rotation rate. On fig.\ref{ramp}(b), the rupture, maximal and dynamical stresses are displayed as a function of packing fraction for a rotation rate ($\overset{.}{\beta}=0.00104rd/s$). The values increase strongly with packing fraction. Interestingly these stress values are strongly related since their ratio stay constant with packing fraction: $ X=\sigma_M/\sigma_r =3.5\pm 0.2 $ (see upper part of fig.(\ref{ramp}b).\\
\begin{figure}[htb]
\begin{center}
\includegraphics[width=0.48\textwidth]{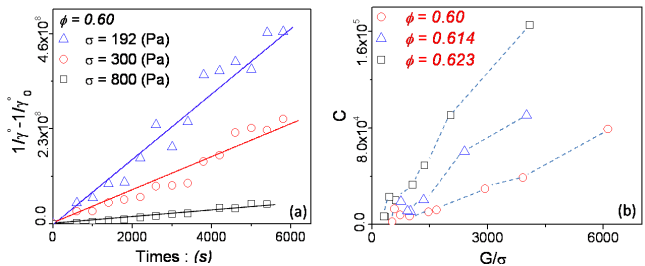}
\caption{Creep experience. (a) plot of  $\overset{.}{\gamma}^{-1}-\overset{.}{\gamma }_{0}^{-1}$ as a function of time for creep experiments performed a constant packing fraction $\phi = 0.6$ for various shear stresses, the straight lines are linear fits : $y=Ct$ . (b) Values of the fitted slopes $C$ as a function of $G(\phi)/\sigma$ for three values of packing fractions.}
\vspace{-0.5cm}
\label{InvShear}
\end{center}
\end{figure}
\textsl{Creep flow~} - This part is concerned with the creep response of the granular packing under constant shear stress. The packing fraction is varied between $0.56$ and $0.625$. The procedure consists of two steps. First, a monotonic loading up to the desired stress value. This initial step is fast with respect to the creep dynamics, typically less than $200s$. We verified that this loading time is much smaller than the inverse of the initial shear rate.
Second, a phase of constant applied shear stress is obtained by a feed-back procedure where stress is maintained at a constant value within a range less than 1 \%. The onset of feed-back defines the initial time $t=0$. If stress leaves the assigned range, a command is sent to the motor to rotate the torsion spring and to adjust the torque accordingly. From time to time, as a consequence of this fast motor rotation, an acceleration of the strain rate is observed, followed by a decay down to the value before the jump. These dynamical phases lasting less than $20s$ were replaced by a linear interpolation of the shear rate so that they do not artificially perturb the subsequent analysis. 
For all the experiments, we observe a slow increase of the deformation $\Delta \gamma(t)$ (see fig.\ref{setup}b). The creep dynamics increases with the applied shear stress and decreases for larger values of the packing fraction. To quantify the creep dynamics, strain rates were computed for a time step $\delta t = 1 s$ and averaged over a time window of $\Delta T= 400s$. On fig.\ref{InvShear}a, we display $\overset{.}{\gamma} ^{-1}-\overset{.}{\gamma }_{0}^{-1}$ as a function of time for a fixed packing fraction at different applied shear stresses; $\overset{.}{\gamma }_{0}$ being the initial shear rate. This representation is a natural choice to probe a long-time logarithmic dynamics (consistent with  $\overset{.}{\gamma}\propto 1/t$). Indeed, we observe a relation of the type: $\overset{.}{\gamma }^{-1}-\overset{.}{\gamma }_{0}^{-1}=Ct$, corresponding to a long time logarithmic creep : $\gamma =\gamma _{0}+\dfrac{\overset{.}{\gamma }_{0}}{C}\ln (1+Ct)$. On fig.\ref{InvShear}b, we represent the values of $C$ extracted from a linear fit as a function of $G(\phi)/\sigma$, for different packing fraction values. We observe a monotonic increase, more prononced at larger packing fraction.\\
\textsl{Rheological model~} - To analyze quantitatively the data, we use a theoretical model introduced by Derec et al.\cite{Derec2001} in the context of complex fluids rheology. We propose here to adapt it to the creep flows of granular packing. The model extends naturally the standard Maxwell visco-elastic rheology. It introduces an internal phenomenological variable $f$ called "fluidity" whose dimension is an inverse time. Physically, fluidity is a rate of stress relaxation. To describe a complex dynamics displaying ageing and rejuvenation, Derec et al. propose that fluidity should be time-dependent and its dynamics described by a simple "à la Landau" phenomenological equation :
\begin{eqnarray}
\partial _{t}\sigma =-f\sigma +G\overset{.}{\gamma }\\
\partial _{t}f=-a f^{2}+r\overset{.}{\gamma }^{2}
\label{EquFluStress}
\end{eqnarray}
The second equation introduces two dimensionless and positive parameters $a$ and $r$. The first term corresponds a fluidity decrease i.e. an ageing process which renders the fluid  more "viscous" with time. We call $a$ the "ageing parameter". The second term, corresponds a fluidity increase (less "viscous") due to shear. We call $r$ the "rejuvenation" parameter. The forms assumed by the ageing and rejuvenation terms are the most simple one can get in an expansion consistent with a non-trivial long-time dynamics (see discussion in \cite{Derec2001}). 
At constant shear stress $\sigma$, one obtains the relation $f \sigma =G\overset{.}{\gamma }$, yielding : $\partial_{t}f=-a(1-(\dfrac{\sigma }{\sigma_D})^{2})f^{2}$, where $\sigma _{D} = G\sqrt{a/r}$ is the dynamical shear corresponding to steady shear rate and steady fluidity. Introducing an equivalent ageing parameter :
\begin{equation}
a_{eq}= a(1-(\dfrac{\sigma }{\sigma_D})^{2})
\label{aeq}
\end{equation}
the solution of this equation is then : $f(t)=\dfrac{f_{0}}{1+a_{eq}ft}$. The shear rate variation is thus : $\overset{.}{\gamma }^{-1}-\overset{.}{\gamma }_{0}^{-1} = \frac{G}{\sigma }{a_{eq}}.t $ which leads to a long time logarithmic creep as observed experimentally. The experimental slopes $C$ of fig. (\ref{InvShear}b) can then be identifed using the relation :
\begin{equation}
{a_{eq}} = C \frac{\sigma }{G}
\label{C}
\end{equation}
For all experiments performed at different stresses and packing fractions, the initial fluidity value $f_{0}=G(\phi)\overset{.}{\gamma_0 }/\sigma$ can be plotted as a function of $\phi$. For all stresses, data collapse onto a quasi-linear curve (see inset of fig.\ref{collapse}); the denser is the packing, the less is the initial fluidity. The linear extrapolation of this curve to $f_{0}=0$ yields a value $\phi_0=0.635\pm 0.002 $, close to random close packing of monodisperse spheres. This can be interpreted as an arrest of the creep dynamics at a packing fraction corresponding to the jamming limit for a random assembly of frictionless spheres \cite{Torquato_2010}. Furthermore, the ageing dynamics can be characterized by computing the equivalent ageing parameter $a_{eq}$ according to relation (\ref{C}). If $a$ and $r$ are independent of shear, $a_{eq}$ should decrease quadratically and reach a zero value at a finite stress corresponding to the dynamical stress $\sigma_D$ according to relation (\ref{aeq}).
Interestingly, the initial fluidity $f_{0}$ sets a time scale which can also be interpreted as an effective viscosity $\eta_0\ = G(\phi) /f_{0}(\phi)$ diverging when approaching the jamming threshold from below : $\eta_0 \propto (\phi_0 - \phi)^{-1}$. Since, in principle, we are well below any thermalized regime where the viscosity concept could apply, this result is quite remarkable. Moreover, we can use this original time scale to rescale the effective ageing parameter and obtain a collapse of all the data. On fig.(\ref{collapse}), we plot $Y=f_{0} a_{eq}$ as a function of the non-dimensionalized stress : $X=\sigma/\sigma_r$. The striking feature is that all data collapse onto a single curve for the entire range of stresses and packing fractions studied. The second important feature is that $a_{eq}$ displays a non monotonic behavior with a minimum value at $\sigma \approx \sigma _{r}$ ($X=1$) corresponding to the onset of the strain rate bursts identified in the stress-ramp experiments. This behavior is clearly the signature of internal granular reorganizations leading to a slowing down of the creep dynamics instead of an increase as one might expect when shear is increased. For values above $X\approx 2$, the creep dynamics increases again before reaching the dynamical stress threshold ($a_{eq}=0$) at $\sigma_D \approx 2.4 \sigma _{r}$. For the smaller values of shear stress, i.e. below $\sigma_r$, the predictions of Derec's model with constant coefficients can still be validated with a dynamical stress $\sigma_D=1.5 \sigma_r $ (see dashed line on fig.\ref{collapse}).\\
	This experimental study shows that down to vanishing low applied shear and up to the yield stress value, internal relaxation processes are present in a granular packing. The logarithmic creep hence observed, was analyzed using a simple visco-elastic model which introduces a time dependent rate of relaxation (the fluidity). The dynamics is viewed as a competition beween intrinsic ageing and shear stress rejuvenation. The model allows a dynamical characterization of the initial packing fluidity which decreases linearly with packing fraction and vanishes at the random close packing limit. Under finite stress, we identified the onset of internal reorganizations, slowing down the creep process and setting the yield stress to higher values. This process could be related to the onset of shear induced anisotropy \cite{Geng_2003} or shear band formation. This internal dynamics is the sign of a peculiar fragility of this type of solid, possibly mediated by thermal activation or by background mechanical noise. It may also be related to the intrinsic nature of the plastic response in amorphous solids \cite{Lemaitre_2009}, which are sometimes described as structurally fragile under finite shear \cite{Hentschel_2010}.\\
We acknowledge the financial of the support CNRS-PEPS program and the ANR \textit{Jamvibe-2010}.

\end{document}